# Challenges in experimental data integration within genome-scale metabolic models

*Pierre-Yves Bourguignon[1,2], Areejit Samal[1,3], François Képès[4], Jürgen Jost[1,5,§], Olivier C. Martin[3,6,§]*

[1]Max Planck Institute for Mathematics in the Sciences, Inselstr. 22, D-04103 Leipzig, Germany
[2]Laboratoire de Physique Statistique, CNRS and Ecole Normale Supérieure, UMR 8550, F-75231 Paris, France
[3]Laboratoire de Physique Théorique et Modèles Statistiques, CNRS and Univ Paris-Sud, UMR 8626, F-91405 Orsay, France
[4]Epigenomics Project, Genopole, CNRS UPS 3201, UniverSud Paris, University of Evry, Genopole Campus 1 - Genavenir 6, Evry, France
[5]The Santa Fe Institute, 1399 Hyde Park Road, Santa Fe, NM 87501, USA
[6]INRA, UMR 0320/UMR 8120 Génétique Végétale, Univ Paris-Sud, F-91190 Gif-sur-Yvette, France
[§]Correspondence should be addressed to: *Jürgen Jost* (jost@mis.mpg.de) or *Olivier C. Martin* (olivier.martin@u-psud.fr)

## Abstract

A report of the meeting "Challenges in experimental data integration within genome-scale metabolic models", Institut Henri Poincaré, Paris, October 10-11 2009, organized by the CNRS-MPG joint program in Systems Biology.

## Meeting Report

The meeting "Challenges in experimental data integration within genome-scale metabolic models" was held at the Institut Henri Poincaré, Université Pierre et Marie Curie, Paris, October 10$^{th}$ and 11$^{th}$, 2009 [1]. It brought together leading international researchers in the field of genome-scale metabolic modelling and enzyme-kinetics modelling. As suggested by the title, the emphasis was on innovative methodologies aimed at taking better advantage of various experimental data types (such as measurements of flux and intra-cellular metabolite concentrations, tracing of isotopomers, mutant growth phenotypes and gene expression datasets). These kinds of data will increasingly empower researchers aiming to characterize metabolism in various biological systems, as well as its evolution. In this report, we outline the most important advances presented at the meeting.

## Model reconstruction and improvement

While the number of fully sequenced genomes continues to grow at an exponential rate, the number of published reconstructions of metabolic models [2] is dramatically lagging behind the sequencing effort. This slow pace of model reconstruction effort was highlighted by both David Fell (Oxford Brookes University, UK) and Costas Maranas (Penn State University, USA) at the meeting. While various automatic procedures have been introduced during this past decade to assist the reconstruction of metabolic models, their output still requires a painstaking curation effort. Fell discussed various kinds of inconsistencies that are prevalent in many existing genome-scale metabolic reconstructions including presence of dead-end metabolites, stoichiometric imbalance of certain reactions and erroneous reaction directionality assignments [3]. He also stressed the need to develop *automated* heuristics for both fast supervised curation of existing models and for the construction of new metabolic models. Instances of such methods were presented by Maranas, who developed with his colleagues novel algorithms including GapFill and GapFind [4] to fill gaps associated with the presence of dead-end metabolites in existing models through proper reaction reversibility assignment and prediction of missing pathways.

While single gene-deletion mutants are considered a prominent source of data for assessing the quality of reconstructed models, datasets including the phenotypes of double gene-deletion mutants appeared recently. Balázs Papp (BRC Szeged, Hungary) presented unpublished results where such a dataset obtained in yeast *S. cerevisiae* from the Charlie Boone Lab [5] was used to curate and improve the existing genome-scale metabolic model. Exhaustive *in silico* enumeration of *all* lethal gene pairs, triplets and quartets using FBA is computationally intractable for any genome-scale metabolic model; instead, Maranas presented a heuristic method based on a bi-level optimization approach which improves considerably the computational time to obtain lethal triplets and quartets (the gain is several orders of magnitude) as candidates for further assessment of the genetic interactions predicted by the model [6].

Tomer Shlomi (Technion University, Israel) also showed that reconstructing a model may involve further challenges, pertaining for instance to the proper account of cellular compartments in absence of prior knowledge of enzyme localization. In particular, he presented a novel algorithm to predict sub-cellular localization of enzymes based on their embedding metabolic network, relying on a parsimony principle which minimizes the number of cross-membrane metabolite transporters [7].

While the static composition of the biomass as a component of a metabolic model is known to influence the results of FBA predictions, little had been proposed to date in order to overcome this limitation of the framework. Maranas presented the GrowMatch [8] method to resolve discrepancies between *in silico* and *in vivo* single mutant growth phenotypes by suitably modifying the static biomass composition under different environmental conditions. Shlomi presented a method, Metabolite-dilution FBA (MD-FBA), which systematically accounts for the growth demand of synthesizing all intermediate metabolites required for balancing their growth dilution, leading to improved metabolic phenotype predictions [9].

Condition-dependent refinements of metabolic models can also be fed by further experimental observations. Recently, $^{13}$C labeling experiments followed by nuclear magnetic resonance (NMR) or mass spectrometry (MS) analysis have generated experimental data for a number of intracellular fluxes and metabolite concentrations [10]. Such experimental data along with Gibbs energies of formation contain valuable thermodynamic information determining the reaction directionalities in genome-scale metabolic models. Matthias Heinemann (ETH Zurich, Switzerland) presented a novel algorithm called Network Embedded Thermodynamic (NET) analysis [11] which systematically assigns reaction directionalities in genome-scale metabolic models using available thermodynamic information.

Another criticism often addressed to FBA pertains to the use of an optimality principle to obtain a single biologically relevant flux distribution. Stefan Schuster (University of Jena, Germany) emphasized that FBA predicts a flux distribution that strictly maximizes biomass yield rather than biomass flux or growth rate. Although, in most situations, maximization of rate and yield give equivalent solutions, Schuster presented interesting examples in *S. cerevisiae* and *Lactobacilli* where the two maximizations are not equivalent. He compared the two cases with the experimentally observed solution corresponding to maximization of rate [12]. In contrast to FBA, the elementary mode or extreme pathway analysis tries to characterize the infinite set of allowable flux distributions in solution space through a finite set of representative flux distributions. However, both elementary mode and extreme pathway analysis [13] cannot be scaled up to analyze genome-scale metabolic networks, and to circumvent these problems, Schuster and colleagues have recently developed the concept of elementary flux patterns [14] closely related to elementary modes which can be applied to genome-scale networks.

**Design features of metabolic networks**

The reconstruction of metabolic networks for several organisms spread across the tree of life and that thrive in diverse habitats has enabled investigations aimed at understanding the role of the environment in determining the structure of metabolic networks of different organisms. Oliver Ebenhöh (University of Aberdeen, UK) presented a simple heuristic based on the principle of forward propagation called network expansion [15] which uses a bipartite graph representation of cellular metabolism to predict the "scope" or synthesizing capability of any metabolite in the investigated network. Using the expansion algorithm and metabolic networks of different organisms in the KEGG database, Ebenhöh and colleagues were able to classify different species as generalists or specialists based on their different carbon utilization spectra [16].

Marie-France Sagot (INRIA, France) presented ongoing work in her group to improve the network expansion algorithm by appropriately differentiating self-regenerating metabolites (usually cofactors) [17] from nutrient metabolites in the starting seed set to predict the minimum set of additional precursor metabolites needed to reach the target metabolites from nutrient metabolites in the environment. She mentioned an interesting application of this algorithm in determining the precursor set that an endosymbiont like *Buchnera aphidicola* receives from its host.

Several studies in the past have been focused towards understanding the relation between structure and function of metabolic networks. However, little is known about the variation in reaction content of the different possible metabolic networks having the same phenotype. Olivier Martin (Univ Paris Sud, France) presented a new method based on Markov Chain Monte Carlo (MCMC) sampling which can be used to uniformly sample the space of metabolic networks with a given phenotype and fixed number of reactions in a global reaction set [18]. Using this method and a hybrid database constructed from KEGG and the *E. coli* metabolic network, Martin and colleagues showed that the *E. coli* network is atypically robust to mutations.

While the investigation of statistically overrepresented motifs in gene regulatory networks has resulted in the identification of qualitative features of the associated dynamics [19], similar attempts in metabolic networks are often deemed hopeless. Andreas Kremling (Max-Planck Institute for Dynamics of Complex Technical Systems, Magdeburg, Germany) presented a successful study [20] where a general scheme underlying catabolic repressions in *E. coli* was identified. Modeling this process allowed him to further characterize qualitatively different regimes.

**Learning quantitative features**

As an alternative to traditional optimization-based predictions, Daniela Calvetti (Case Western University, USA) presented a probabilistic extension of both kinetic and steady state models of metabolism that she introduced with her colleague E. Somersalo [21]. Relying on Bayesian induction, their approach aims to account for the remaining uncertainty after experimental data have been analyzed by outputting posterior distributions rather than sets of achievable states. Appealing features of their framework in comparison to linear programming approaches include the absence of a hypothesized objective function, the tolerance to model mis-specifications, as well as the assessment of the probability of a particular solution. This latter feature is of particular interest when multiple experimental conditions are to be compared. Various applications of this framework to the assessment of candidate mechanisms underpinning various metabolic changes were also presented.

Wolfram Liebermeister (Humboldt University, Berlin, Germany) presented various methods leveraging such mathematical theories to integrate experimental data within metabolic models. He provided the audience with a thorough review of the methods he developed with his colleagues to induce quantitative relationships between enzyme levels, metabolite concentrations and metabolic fluxes, while properly accounting for physical laws and allosteric regulation [22-23]. Emphasis was put on the thermodynamic relevance of kinetic laws, as well as on the importance of accounting for the uncertainty pertaining to their parameters. Besides theoretical considerations, he also mentioned how computationally tractable inferences of kinetic laws can be achieved.

**Human metabolism**

Although the detailed modelling of human metabolism was initiated almost ten years ago, to date it has been restricted to specific cell-types and organelles. In parallel, comprehensive datasets of the genes involved and biochemical activities in human cells have been gathered, allowing Duarte and colleagues to publish the first global map of human metabolism in 2007 [24]. Building upon this wealth of knowledge, Eytan Ruppin (Tel Aviv University, Israel) undertook the reconstruction of tissue-specific pathways using gene expression data, and presented at this meeting both the methods [25] that his team developed and some of the applications of their use. On the methodological side, traditional reconstruction techniques using the FBA framework needed in-depth adaptations: the fundamental ingredients, namely the specification of the medium and the objective function, are indeed unknown in this particular setting. Using the agreement between expression data and flux values as an objective function, they developed a Mixed Integer Linear Programming approach to meet the requirements of their project. This approach was further validated, and even post-transcriptional regulation could be investigated in their framework. An application of this framework for predicting biomarkers of genetic errors of metabolism was also presented [26]. Finally, Ruppin described another approach aimed at reconstructing tissue-specific models of metabolism by successively removing dispensable reactions and then activating other reactions known to occur in the tissue of interest. An application to the reconstruction of a model of liver metabolism was used to illustrate the method

Kiran Patil (Technical University of Denmark, Denmark) tackled the challenge of modeling several other metabolic processes in humans. He specifically investigated the metabolic and regulatory underpinnings of diabetes, combining the knowledge on regulatory and metabolic mechanisms to pinpoint biomarkers of diabetes with the help of several case-studies pertaining to this particular disease. An analysis of the enrichment in binding sites of transcription factors in upstream regions of the enzymatic genes relevant to this study allowed him to uncover the potential of various transcription factors as drug targets [27].

**Acknowledgement**

We thank Antje Vandenberg, Corine Legrand, Florence Lajoinie, Heiko Schinke, Sylvie Dubois, Saskia Gutzschebauch and Katrin Scholz for their help, administrative support and making the meeting a success.